\documentclass[twocolumn,showpacs,preprintnumbers,amsmath,amssymb]{revtex4}
\date{May 2009}
\usepackage{epsfig}
\usepackage{latexsym}
\usepackage{amssymb}
\usepackage{booktabs}
\usepackage{graphicx}
\newcommand{\be}{\begin{equation}}
\newcommand{\ee}{\end{equation}}
\newcommand{\ba}{\begin{eqnarray}}
\newcommand{\ea}{\end{eqnarray}}
\newcommand{\bi}{\begin{itemize}}
\newcommand{\ei}{\end{itemize}}
\newcommand{\tr}{{\rm Tr\,}}

\newcommand{\quarter}{{\textstyle\frac{1}{4}}}

\newcommand{\<}{\langle}
\renewcommand{\>}{\rangle}
\newcommand{\eq}{Eq.~}

\newcommand{\fig}{Fig.~}

\newcommand{\la}{\label}

\newcommand{\txts}{\textstyle}

\begin{document}
\preprint{MIT-CTP 4041}
\title{High-Precision Thermodynamics and Hagedorn Density of States}

\author{Harvey~B.~Meyer}
\affiliation{Center for Theoretical Physics\\
      Massachusetts Institute of Technology\\
    Cambridge, MA 02139, U.S.A.}

\date{\today}

\begin{abstract}
We compute the entropy density of the confined phase 
of QCD without quarks on the lattice to very high accuracy.
The results are compared to the entropy density
of free glueballs, where we include
all the known glueball states below the two-particle threshold. 
We find that an excellent, parameter-free description of
the entropy density between $0.7T_c$ and $T_c$
is obtained by extending the spectrum
with the exponential spectrum of the closed bosonic string.
\end{abstract}

\maketitle

\section{Introduction}
The phase diagram of quantum chromodynamics (QCD) 
is being actively studied in heavy ion collision experiments
as well as  theoretically. A form of matter with 
remarkable properties~\cite{Muller:2008zzm}
has been observed in the Relativistic Heavy Ion Collider (RHIC)
experiments~\cite{Arsene:2004fa,Back:2004je,Adcox:2004mh,Adams:2005dq}. 
It appears to be a strongly coupled
plasma of quarks and gluons (QGP), but no consensus on 
a physical picture 
that accounts for both equilibrium and non-equilibrium properties
has been reached yet.
On the other hand, below the short interval of temperatures
where the transition from the confined phase to the 
QGP takes place~\cite{Aoki:2006br,Aoki:2009sc,Cheng:2007jq,Bazavov:2009zn},
it is widely believed that the most prominent degrees of freedom are the 
ordinary hadrons. From this point of view, 
the zeroth order approximation to the properties
of the system is to treat the hadrons as infinitely
narrow and non-interacting. We will refer to this approximation
as the hadron resonance gas model (HRG).
The HRG predictions were compared 
with lattice QCD thermodynamics data 
in~\cite{Karsch:2003vd,Cheng:2007jq}, and 
lately they have been used to extrapolate certain results 
to zero temperature~\cite{Bazavov:2009zn}.
The HRG is also the basis of the statistical model currently applied
to the analysis of hadron yields in 
heavy ion collisions~\cite{Andronic:2008gu},
and recently the transport properties of a relativistic hadron gas 
have been studied in detail~\cite{Demir:2008tr}.

Since any heavy ion reaction ends up in the low-temperature phase
of QCD, it is important to understand its properties in detail
in order to extract those of the high-temperature phase
with minimal uncertainty. In this Letter we study whether 
the HRG model works in the absence of quarks, in other words 
in the pure SU($N=3$) gauge theory, where the low-lying states
are glueballs. There are reasons to believe that 
if the HRG model is to work at any
quark content of QCD, it is in the zero-flavor case.
Firstly, the mass gap in SU(3) gauge theory is very large, 
$M_0/T_c\simeq 5.3$. As we shall see, 
the thermodynamic properties up to quite close to $T_c$ 
are dominated  by the states below the two-particle threshold, 
which are exactly stable. Furthermore, because of their large mass, 
neglecting their thermal width should be a good approximation.
Secondly, the scattering amplitudes between glueballs are 
parametrically $1/N^2$  suppressed 
while those between mesons are only $1/N$
suppressed~\cite{Witten:1979kh}. 
This means that the glueballs should be free to a 
better approximation than the hadrons of realistic QCD.

An additional motivation to study the thermodynamics of the 
confined phase of SU(3) gauge theory is that it is a parameter-free
theory, simplifying the interpretation of its properties.
Its spectrum is known quite accurately up to the two-particle 
threshold~\cite{Meyer:2004gx,Chen:2005mg}. By contrast,
in full QCD calculations, lattice data calculated 
at pion masses larger than in Nature
are often compared out of necessity to the HRG model based on the 
experimental spectrum~\cite{Cheng:2007jq,Bazavov:2009zn}.
Finally, calculations in the pure gauge theory
are at least two orders of magnitude faster, which allows us to 
reach a high level of control of statistical and systematic errors;
in particular, we are able to  perform calculations 
in very large volumes.

\section{Lattice calculation}
We use Monte-Carlo simulations of the Wilson action 
$S_{\rm g}= \frac{1}{g_0^2} \sum_{x,\mu,\nu} \tr\{1-P_{\mu\nu}(x)\} $
for SU(3) gauge theory~\cite{Wilson:1974sk}, where $P_{\mu\nu}$ is the plaquette.
The lattice spacing is related to the bare coupling through 
$g_0^2\sim 1/\log(1/a\Lambda)$.
We calculate the thermal expectation value of
$\theta\equiv T_{\mu\mu}$, the (anomalous) 
trace of the energy-momentum tensor $T_{\mu\nu}$, and 
of $\theta_{00}\equiv T_{00}-\quarter \theta$. 
In the thermodynamic limit,
\be
 Ts = e+p = {\txts\frac{4}{3}} \<\theta_{00}\>_T,~~~~
e-3p= \<\theta\>_T - \<\theta\>_0.
\ee
Here $e,p,s$ are respectively the energy density, pressure and entropy density.
The operator 
$\theta_{00}=\frac{1}{2}(-{\bf E}^a\cdot{\bf E}^a+{\bf B}^a\cdot{\bf B}^a)$
requires no subtraction, because its vacuum expectation value vanishes.
The choice of of $\theta_{00}$ and $\theta$ as 
independent linear combinations is convenient
because they both renormalize multiplicatively.
We use the `HYP-clover' discretization of the energy-momentum tensor 
introduced in~\cite{Meyer:2007tm,Meyer:2007ed}.
The normalization of the $\theta_{00}$ operator differs from its
naive value by a factor that we parametrize as $Z(g_0)\chi(g_0)$. 
The factor $Z(g_0)$ is taken from~\cite{Meyer:2007ic} and 
rests on the results of~\cite{Engels:1999tk}; its accuracy is about one percent.
The factor $\chi(g_0)$ is obtained by calibrating
our discretization to the `bare plaquette' discretization in the 
deconfined phase at $N_t=6$~\cite{Meyer:2007tm}. 
We find, for $6/g_0^2$ between 5.90 and 6.41,
$\chi(g_0)=0.1306\cdot(6/g_0^2)-0.1865$ with an accuracy of half a percent.
For the lattice beta-function that renormalizes $\theta$, we use 
the parametrization~\cite{Durr:2006ky} of the data in~\cite{Necco:2001xg}
and the same calibration method. 
\begin{figure}
\centerline{\includegraphics[width=6.5 cm,angle=-90]{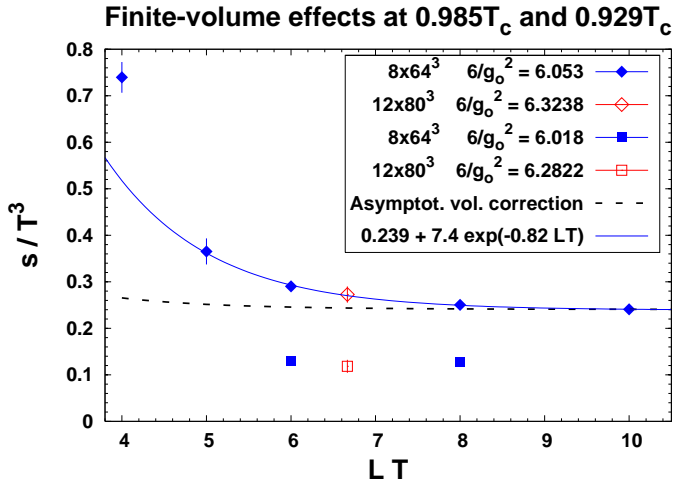}}
\caption{Finite volume effects on the entropy density
 close to the deconfining temperature $T_c$.}
\label{fig:fv}
\end{figure}

Our results for the entropy density from $N_t=8$ and $N_t=12$ 
simulations are shown on \fig(\ref{fig:epp}). The displayed error
bars do not contain the uncertainty on the normalization factor,
which is much smaller and would introduce correlation between the points. This factor 
varies by only $7\%$ over the displayed interval and so 
to a first approximation amounts to an overall normalization of the curve.
Our data is about five times statistically more accurate than that of 
previous thermodynamic studies~\cite{Boyd:1996bx,Namekawa:2001ih}, 
which were primarily focused on the deconfined phase.
Just as importantly, we kept the finite-spatial-volume effects under good
control, in particular very close to $T_c$. 

Figure (\ref{fig:fv})
shows the size of finite-volume effects. For instance, at $0.985T_c$
the conventional choice $LT=4$ leads to an overestimate of the entropy 
density by a factor three. The fact that the $N_t=12$ data fall on the 
same smooth curve as the $N_t=8$ is strong evidence that discretization
errors are small. We parametrize the volume dependence empirically
by a $A+Be^{-cLT}$ curve, and use it to convert the $N_t=12$ data to $LT=8$. 
At $0.929T_c$, there is no statistically significant
difference between $LT=6$ and 8 and we do not apply any correction.
It is the corrected $N_t=12$ data that is then displayed on \fig(\ref{fig:epp}).

In~\cite{Meyer:2009kn}, formulas for the leading finite-volume effects 
on the thermodynamic potentials were derived in terms of the energy gap 
of the theory defined on a $(1/T)\times L\times L$ spatial hypertorus.
Close to $T_c$, this gap corresponds to the mass of the ground state
flux loop winding around the cycle of length $1/T$. 
If $\delta s(T,L) \equiv s(T,\infty)-s(T,L)$, the formula then reads
\be
\delta s(T,L) = \frac{e^{-m(T) L}}{2\pi L}
\left[  m^2(T) + {\txts\frac{3}{2}}T\partial_T m^2(T)\right]\,.
\la{eq:s}
\ee
Using the calculation of $m(T)$ described in the next section,
the predicted asymptotic approach to the
infinite-volume entropy density 
for $0.985T_c$ is displayed on \fig(\ref{fig:fv}). 
While the sign is correct, 
the magnitude of the finite-volume effects 
is not reproduced for $LT\leq 8$. We conclude that the 
asymptotic approach to infinite volume sets in for
very large values of $LT$. Since $m(T)L$ is only about $4$ when $LT=6$, 
it is not implausible that flux-loop states with high multiplicity
dominate the finite-volume effects at that box size.

\begin{figure}
\centerline{\includegraphics[width=6.5 cm,angle=-90]{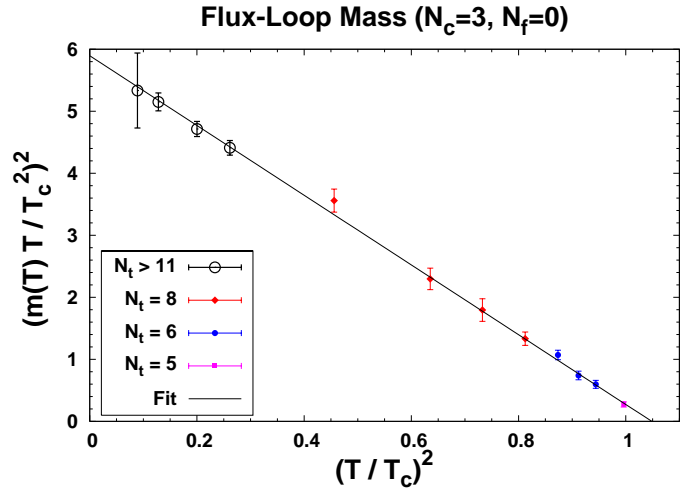}}
\caption{The mass of the temporal flux loop as calculated from 
Polyakov loop correlators, and the fit (\ref{eq:II}).
The $N_t>11$ data are from~\cite{Meyer:2004vr}, the 
$N_t=5$ data from~\cite{Lucini:2005vg}.}
\label{fig:tmass4}
\end{figure}
Next we obtain the correlation length $\xi(T)$ of the order parameter
for the deconfining phase transition, the Polyakov loop.
The method consists in computing the two-point function
of zero-momentum operators, designed to have large overlaps 
with the ground state flux loop, along a spatial direction.
We fit the lattice data for $m(T)\equiv1/\xi(T)$
displayed on \fig(\ref{fig:tmass4}) with the formula
\be
{\txts\left(\frac{m(T)T}{T_c^2}\right)}^2 = a_0 
- a_1 {\txts\left(\frac{T}{T_c}\right)}^2
 - a_2 {\txts\left(\frac{T}{T_c}\right)}^4
\ee
and find, either fitting $a_2$ or setting it to zero, 
\ba
&& a_0= 5.76(15),~ a_1 = 4.97(65), ~ a_2 = 0.55(54)
\la{eq:I}
\\
&& a_0= 5.90(9),~~ a_1 = 5.62(10),~~ a_2 = 0
\la{eq:II}
\ea
with in both cases a $\chi^2/$dof of about 0.3.
We remark that the $a_i$ are not far from 
the Nambu-Goto string~\cite{Arvis:1983fp} values
$a_1=\frac{2\pi}{3}\frac{\sigma}{T_c^2}= 5.02(5)$~\cite{Lucini:2005vg}
and $a_2=0$ ($\sigma$ is the  tension of the confining string).
We extract the `Hagedorn' temperature, defined 
as in~\cite{Bringoltz:2005xx} by  $m(T_h)=0$,
from the second fit,
\be
T_h/T_c = 1.024(3).
\la{eq:Th}
\ee
This extraction amounts to assuming mean-field exponents
near $T_h$ (it is not clear which universality class should be 
used~\cite{Yaffe:1982qf}). The result is stable if 
the fit interval is varied, and also if $a_2$ is fitted with 
$a_0$ and $a_1$ constrained to the
known values of $(\sigma/T_c^2)^2$ and $\frac{2\pi}{3}\frac{\sigma}{T_c^2}$.

As a check on the normalization of the operators
$\theta_{00}$ and $\theta$,
we calculate the latent heat in two different ways. The latent  heat 
is the jump in energy density at $T_c$. Since the pressure
is continuous, we obtain it instead from the discontinuity
in entropy density or the `conformality measure' $e-3p$.
We obtain $s$ and $e-3p$ on either side of $T_c$
by extrapolating $LT=10$  data from the confined (deconfined)
phase towards $T_c$. The result is
\be
\frac{\Delta s}{T_c^3} = 1.45(5)(5),~~
\frac{\Delta(e-3p)}{T_c^4} =  1.39(4)(5),
\ee
where the first error is statistical and the second comes from 
the uncertainty in the extrapolation (taken to be the difference
between a linear and quadratic fit).
The compatibility between these two estimates of $L_h/T_c^4$
is strong evidence that we control the normalization of 
our operators. They are in good agreement with previous 
calculations of the latent heat performed 
on coarser lattices 
\cite{Beinlich:1996xg,Lucini:2005vg}.
We have also verified more generally that the thermodynamic identity
$T\partial_T(s/T^3)=(1/T^3)\partial_T(e-3p)$ 
is satisfied within statistical errors.

\section{Interpretation}
In infinite volume the pressure associated with a single non-interacting,
relativistic particle species 
of mass $M$ with $n_\sigma$ polarization states reads
\be
p = \frac{n_\sigma}{2\pi^2}M^2\,T^2\sum_{n=1}^\infty  \frac{1}{n^2} K_2(nM/T)
\ee
where $K_2$ is a modified Bessel function.
By linearity, the knowledge of the glueball spectrum 
leads to a simple prediction for the pressure and entropy density 
$s=\frac{\partial p}{\partial T}$,
which is expected to become exact in the large-$N$ limit.
Since only the low-lying spectrum of glueballs is known, it 
is useful to consider how the density of states might  be extended above
the two-particle threshold $2M_0$, where $M_0$ is the mass of the 
lightest (scalar) glueball.
The asymptotic closed bosonic string density of states in four dimensions 
is given by~\cite{Zwiebach:2004tj}
\be
\rho(M) = \frac{(2\pi)^3}{27\,T_h} \left(\frac{T_h}{M}\right)^4 e^{M/T_h}.
\la{eq:hagedorn}
\ee
In the string theory, the Hagedorn temperature $T_h$ is related to the 
string tension, $T_h^2=\frac{3\sigma}{2\pi}$, corresponding to 
$T_h/T_c=1.069(5)$~\cite{Lucini:2003zr}. Below we use 
this value as an alternative to the more direct determination (\ref{eq:Th}).

\begin{figure}
\centerline{\includegraphics[width=6.5 cm,angle=-90]{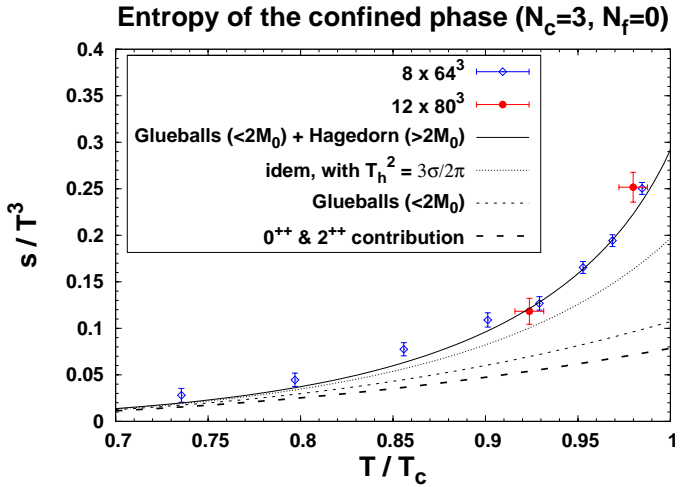}}
\caption{The entropy density in units of $T^3$ for $LT=8$.
We applied a (modest) volume-correction to the $N_t=12$ data.}
\label{fig:epp}
\end{figure}

On \fig\ref{fig:epp}, we show the entropy contribution of the glueballs 
lying below the two-particle threshold $2M_0$. The curve is just about consistent
with the smallest temperature lattice data point, but clearly fails to 
reproduce the strong increase in entropy density as $T\to T_c$.
The figure also illustrates that the two lowest-lying states, the 
scalar and tensor glueballs, account for about three quarters 
of the stable glueballs' contribution.
We have used the continuum-extrapolated 
lattice spectrum~\cite{Meyer:2004gx,Meyer:2008tr}.

Adding the Hagedorn spectrum contribution, \eq(\ref{eq:hagedorn})
with $T_h$ given by \eq(\ref{eq:Th}),
leads to the solid curve on \fig\ref{fig:epp}. It describes the direct 
calculation of the entropy density surprisingly well, particularly 
close to $T_c$. The curve tends to underestimate
somewhat the entropy density at the lower temperatures. 
This is likely to be a cutoff effect. Indeed, at fixed $N_t$ 
lower temperatures correspond to a coarser lattice spacing,
and the scalar glueball mass in physical units is known 
to be smaller on coarse lattices with the Wilson action~\cite{Lucini:2001ej}.
If we use the stable glueball spectrum calculated at $g_0^2=1$
instead of the continuum spectrum, the agreement of the non-interacting 
glueball + Hagedorn spectrum with the lattice data at the lower
four temperatures is again excellent. 
This difference provides an estimate for the size of lattice effects.

To summarize, we have computed to high accuracy 
the entropy of the confined phase of 
QCD without quarks. The low-lying states
of the theory are therefore bound states 
called glueballs, and their spectrum is well 
determined~\cite{Meyer:2004gx,Chen:2005mg}.
If the size $N$ of the gauge group is increased, the interactions
of the glueballs are expected to be suppressed~\cite{Witten:1979kh}. 
To what extent the glueballs really are weakly interacting at $N=3$
is not known precisely. Some evidence for the smallness of their low-energy  
interactions was found some time ago         
by looking at the finite-volume effects on their masses~\cite{Meyer:2004vr}. 
But it seems unlikely that glueballs well above the two-particle    
threshold would have a small decay width.                          
We have nevertheless compared the entropy density data to 
the entropy density of a gas of non-interacting glueballs.
While restricting the spectral sum to the stable glueballs
leads to an underestimate by at least  a factor two of the 
entropy density near $T_c$, extending the spectral sum
with an exponential spectrum $\rho(M)\sim \exp(M/T_h)$, 
suggested long ago by Hagedorn~\cite{Hagedorn:1965st}, leads to a prediction in excellent 
agreement with the lattice data for the entropy density (\fig\ref{fig:epp}). 
This is remarkable, 
since the analytic form of the asymptotic spectrum 
is completely predicted by free bosonic string theory, 
including its overall normalization (\eq\ref{eq:hagedorn}). Therefore, 
since we also separately computed the temperature (identified with $T_h$)
where the flux loop mass vanishes,
no parameter was fitted in the comparison with the thermodynamic data.
By contrast, the entropy density is not nearly as well described 
if the Nambu-Goto value of $T_h$ is used, see \fig(\ref{fig:epp}).

The success of the non-interacting string density of states
in reproducing the entropy density
suggests that once the Hagedorn temperature has been determined 
directly from the divergence of the flux-loop correlation length, 
the residual effects of interactions on the thermodynamic potentials are small. 
It may be that thermodynamic properties in general
are not strongly influenced by interactions when a large number 
of states are contributing. A well-known example  
is provided by the ${\cal N}=4$ super-Yang-Mills theory, 
whose entropy density at very strong coupling                     
is only reduced by a factor 3/4 with respect to the   
free theory~\cite{Gubser:1996de}.                                   
In this interpretation,
the main effect of interactions among glueballs on thermodynamic properties
is to slightly shift the value of the Hagedorn
temperature $T_h$ with respect to its free-string value.
A possible mechanism is that the string tension
that effectively determines $T_h$ is an in-medium string tension
that is $\sim8\%$ lower than at $T=0$.

Returning to full QCD, our results lend support to the idea 
that the hadron resonance gas model can largely account for the 
thermodynamic properties of the low-temperature phase.
Whether the open string density of states 
reproduces the entropy calculated on the lattice 
can also be tested at quark masses not necessarily as light as in Nature
using a simple open string model~\cite{Selem:2006nd}.

\acknowledgments{
I thank B. Zwiebach for a discussion on the bosonic string 
density of states. The simulations were done on the Blue Gene L rack
and the desktop machines of the
Laboratory for Nuclear Science at M.I.T. 
This work was supported in part by
funds provided by the U.S. Department of Energy 
under cooperative research agreement DE-FG02-94ER40818.
}

\bibliography{/afs/lns.mit.edu/user/meyerh/BIBLIO/viscobib}

\begin{thebibliography}{38}
\expandafter\ifx\csname natexlab\endcsname\relax\def\natexlab#1{#1}\fi
\expandafter\ifx\csname bibnamefont\endcsname\relax
  \def\bibnamefont#1{#1}\fi
\expandafter\ifx\csname bibfnamefont\endcsname\relax
  \def\bibfnamefont#1{#1}\fi
\expandafter\ifx\csname citenamefont\endcsname\relax
  \def\citenamefont#1{#1}\fi
\expandafter\ifx\csname url\endcsname\relax
  \def\url#1{\texttt{#1}}\fi
\expandafter\ifx\csname urlprefix\endcsname\relax\def\urlprefix{URL }\fi
\providecommand{\bibinfo}[2]{#2}
\providecommand{\eprint}[2][]{\url{#2}}

\bibitem[{\citenamefont{Muller}(2008)}]{Muller:2008zzm}
\bibinfo{author}{\bibfnamefont{B.}~\bibnamefont{Mueller}},
  \bibinfo{journal}{Prog. Theor. Phys. Suppl.} \textbf{\bibinfo{volume}{174}},
  \bibinfo{pages}{103} (\bibinfo{year}{2008}).

\bibitem[{\citenamefont{Arsene et~al.}(2005)}]{Arsene:2004fa}
\bibinfo{author}{\bibfnamefont{I.}~\bibnamefont{Arsene}} \bibnamefont{et~al.}
  (\bibinfo{collaboration}{BRAHMS}), \bibinfo{journal}{Nucl. Phys.}
  \textbf{\bibinfo{volume}{A757}}, \bibinfo{pages}{1} (\bibinfo{year}{2005}),
  \eprint{nucl-ex/0410020}.

\bibitem[{\citenamefont{Back et~al.}(2005)}]{Back:2004je}
\bibinfo{author}{\bibfnamefont{B.~B.} \bibnamefont{Back}} \bibnamefont{et~al.},
  \bibinfo{journal}{Nucl. Phys.} \textbf{\bibinfo{volume}{A757}},
  \bibinfo{pages}{28} (\bibinfo{year}{2005}), \eprint{nucl-ex/0410022}.

\bibitem[{\citenamefont{Adcox et~al.}(2005)}]{Adcox:2004mh}
\bibinfo{author}{\bibfnamefont{K.}~\bibnamefont{Adcox}} \bibnamefont{et~al.}
  (\bibinfo{collaboration}{PHENIX}), \bibinfo{journal}{Nucl. Phys.}
  \textbf{\bibinfo{volume}{A757}}, \bibinfo{pages}{184} (\bibinfo{year}{2005}),
  \eprint{nucl-ex/0410003}.

\bibitem[{\citenamefont{Adams et~al.}(2005)}]{Adams:2005dq}
\bibinfo{author}{\bibfnamefont{J.}~\bibnamefont{Adams}} \bibnamefont{et~al.}
  (\bibinfo{collaboration}{STAR}), \bibinfo{journal}{Nucl. Phys.}
  \textbf{\bibinfo{volume}{A757}}, \bibinfo{pages}{102} (\bibinfo{year}{2005}),
  \eprint{nucl-ex/0501009}.

\bibitem[{\citenamefont{Cheng et~al.}(2008)}]{Cheng:2007jq}
\bibinfo{author}{\bibfnamefont{M.}~\bibnamefont{Cheng}} \bibnamefont{et~al.},
  \bibinfo{journal}{Phys. Rev.} \textbf{\bibinfo{volume}{D77}},
  \bibinfo{pages}{014511} (\bibinfo{year}{2008}), \eprint{0710.0354}.

\bibitem[{\citenamefont{Bazavov et~al.}(2009)}]{Bazavov:2009zn}
\bibinfo{author}{\bibfnamefont{A.}~\bibnamefont{Bazavov}} \bibnamefont{et~al.}
  (\bibinfo{year}{2009}), \eprint{0903.4379}.

\bibitem[{\citenamefont{Aoki et~al.}(2006)\citenamefont{Aoki, Fodor, Katz, and
  Szabo}}]{Aoki:2006br}
\bibinfo{author}{\bibfnamefont{Y.}~\bibnamefont{Aoki}},
  \bibinfo{author}{\bibfnamefont{Z.}~\bibnamefont{Fodor}},
  \bibinfo{author}{\bibfnamefont{S.~D.} \bibnamefont{Katz}}, \bibnamefont{and}
  \bibinfo{author}{\bibfnamefont{K.~K.} \bibnamefont{Szabo}},
  \bibinfo{journal}{Phys. Lett.} \textbf{\bibinfo{volume}{B643}},
  \bibinfo{pages}{46} (\bibinfo{year}{2006}), \eprint{hep-lat/0609068}.

\bibitem[{\citenamefont{Aoki et~al.}(2009)}]{Aoki:2009sc}
\bibinfo{author}{\bibfnamefont{Y.}~\bibnamefont{Aoki}} \bibnamefont{et~al.}
  (\bibinfo{year}{2009}), \eprint{0903.4155}.

\bibitem[{\citenamefont{Karsch et~al.}(2003)\citenamefont{Karsch, Redlich, and
  Tawfik}}]{Karsch:2003vd}
\bibinfo{author}{\bibfnamefont{F.}~\bibnamefont{Karsch}},
  \bibinfo{author}{\bibfnamefont{K.}~\bibnamefont{Redlich}}, \bibnamefont{and}
  \bibinfo{author}{\bibfnamefont{A.}~\bibnamefont{Tawfik}},
  \bibinfo{journal}{Eur. Phys. J.} \textbf{\bibinfo{volume}{C29}},
  \bibinfo{pages}{549} (\bibinfo{year}{2003}), \eprint{hep-ph/0303108}.

\bibitem[{\citenamefont{Andronic et~al.}(2009)\citenamefont{Andronic,
  Braun-Munzinger, and Stachel}}]{Andronic:2008gu}
\bibinfo{author}{\bibfnamefont{A.}~\bibnamefont{Andronic}},
  \bibinfo{author}{\bibfnamefont{P.}~\bibnamefont{Braun-Munzinger}},
  \bibnamefont{and} \bibinfo{author}{\bibfnamefont{J.}~\bibnamefont{Stachel}},
  \bibinfo{journal}{Phys. Lett.} \textbf{\bibinfo{volume}{B673}},
  \bibinfo{pages}{142} (\bibinfo{year}{2009}), \eprint{0812.1186}.

\bibitem[{\citenamefont{Demir and Bass}(2008)}]{Demir:2008tr}
\bibinfo{author}{\bibfnamefont{N.}~\bibnamefont{Demir}} \bibnamefont{and}
  \bibinfo{author}{\bibfnamefont{S.~A.} \bibnamefont{Bass}}
  (\bibinfo{year}{2008}), \eprint{0812.2422}.

\bibitem[{\citenamefont{Witten}(1979)}]{Witten:1979kh}
\bibinfo{author}{\bibfnamefont{E.}~\bibnamefont{Witten}},
  \bibinfo{journal}{Nucl. Phys.} \textbf{\bibinfo{volume}{B160}},
  \bibinfo{pages}{57} (\bibinfo{year}{1979}).

\bibitem[{\citenamefont{Meyer}(2004)}]{Meyer:2004gx}
\bibinfo{author}{\bibfnamefont{H.~B.} \bibnamefont{Meyer}}
  (\bibinfo{year}{2004}), \eprint{hep-lat/0508002}.

\bibitem[{\citenamefont{Chen et~al.}(2006)}]{Chen:2005mg}
\bibinfo{author}{\bibfnamefont{Y.}~\bibnamefont{Chen}} \bibnamefont{et~al.},
  \bibinfo{journal}{Phys. Rev.} \textbf{\bibinfo{volume}{D73}},
  \bibinfo{pages}{014516} (\bibinfo{year}{2006}), \eprint{hep-lat/0510074}.

\bibitem[{\citenamefont{Wilson}(1974)}]{Wilson:1974sk}
\bibinfo{author}{\bibfnamefont{K.~G.} \bibnamefont{Wilson}},
  \bibinfo{journal}{Phys. Rev.} \textbf{\bibinfo{volume}{D10}},
  \bibinfo{pages}{2445} (\bibinfo{year}{1974}).

\bibitem[{\citenamefont{Meyer and Negele}(2008)}]{Meyer:2007tm}
\bibinfo{author}{\bibfnamefont{H.~B.} \bibnamefont{Meyer}} \bibnamefont{and}
  \bibinfo{author}{\bibfnamefont{J.~W.} \bibnamefont{Negele}},
  \bibinfo{journal}{Phys. Rev.} \textbf{\bibinfo{volume}{D77}},
  \bibinfo{pages}{037501} (\bibinfo{year}{2008}), \eprint{0707.3225}.

\bibitem[{\citenamefont{Meyer and Negele}(2007)}]{Meyer:2007ed}
\bibinfo{author}{\bibfnamefont{H.~B.} \bibnamefont{Meyer}} \bibnamefont{and}
  \bibinfo{author}{\bibfnamefont{J.~W.} \bibnamefont{Negele}},
  \bibinfo{journal}{PoS} \textbf{\bibinfo{volume}{LATTICE2007}},
  \bibinfo{pages}{154} (\bibinfo{year}{2007}), \eprint{0710.0019}.

\bibitem[{\citenamefont{Meyer}(2007)}]{Meyer:2007ic}
\bibinfo{author}{\bibfnamefont{H.~B.} \bibnamefont{Meyer}},
  \bibinfo{journal}{Phys. Rev.} \textbf{\bibinfo{volume}{D76}},
  \bibinfo{pages}{101701} (\bibinfo{year}{2007}), \eprint{0704.1801}.

\bibitem[{\citenamefont{Engels et~al.}(2000)\citenamefont{Engels, Karsch, and
  Scheideler}}]{Engels:1999tk}
\bibinfo{author}{\bibfnamefont{J.}~\bibnamefont{Engels}},
  \bibinfo{author}{\bibfnamefont{F.}~\bibnamefont{Karsch}}, \bibnamefont{and}
  \bibinfo{author}{\bibfnamefont{T.}~\bibnamefont{Scheideler}},
  \bibinfo{journal}{Nucl. Phys.} \textbf{\bibinfo{volume}{B564}},
  \bibinfo{pages}{303} (\bibinfo{year}{2000}), \eprint{hep-lat/9905002}.

\bibitem[{\citenamefont{Duerr et~al.}(2007)\citenamefont{Duerr, Fodor,
  Hoelbling, and Kurth}}]{Durr:2006ky}
\bibinfo{author}{\bibfnamefont{S.}~\bibnamefont{Duerr}},
  \bibinfo{author}{\bibfnamefont{Z.}~\bibnamefont{Fodor}},
  \bibinfo{author}{\bibfnamefont{C.}~\bibnamefont{Hoelbling}},
  \bibnamefont{and} \bibinfo{author}{\bibfnamefont{T.}~\bibnamefont{Kurth}},
  \bibinfo{journal}{JHEP} \textbf{\bibinfo{volume}{04}}, \bibinfo{pages}{055}
  (\bibinfo{year}{2007}), \eprint{hep-lat/0612021}.

\bibitem[{\citenamefont{Necco and Sommer}(2002)}]{Necco:2001xg}
\bibinfo{author}{\bibfnamefont{S.}~\bibnamefont{Necco}} \bibnamefont{and}
  \bibinfo{author}{\bibfnamefont{R.}~\bibnamefont{Sommer}},
  \bibinfo{journal}{Nucl. Phys.} \textbf{\bibinfo{volume}{B622}},
  \bibinfo{pages}{328} (\bibinfo{year}{2002}), \eprint{hep-lat/0108008}.

\bibitem[{\citenamefont{Boyd et~al.}(1996)}]{Boyd:1996bx}
\bibinfo{author}{\bibfnamefont{G.}~\bibnamefont{Boyd}} \bibnamefont{et~al.},
  \bibinfo{journal}{Nucl. Phys.} \textbf{\bibinfo{volume}{B469}},
  \bibinfo{pages}{419} (\bibinfo{year}{1996}), \eprint{hep-lat/9602007}.

\bibitem[{\citenamefont{Namekawa et~al.}(2001)}]{Namekawa:2001ih}
\bibinfo{author}{\bibfnamefont{Y.}~\bibnamefont{Namekawa}} \bibnamefont{et~al.}
  (\bibinfo{collaboration}{CP-PACS}), \bibinfo{journal}{Phys. Rev.}
  \textbf{\bibinfo{volume}{D64}}, \bibinfo{pages}{074507}
  (\bibinfo{year}{2001}), \eprint{hep-lat/0105012}.

\bibitem[{\citenamefont{Meyer}(2009{\natexlab{a}})}]{Meyer:2009kn}
\bibinfo{author}{\bibfnamefont{H.~B.} \bibnamefont{Meyer}}
  (\bibinfo{year}{2009}{\natexlab{a}}), \eprint{0905.1663}.

\bibitem[{\citenamefont{Meyer}(2005)}]{Meyer:2004vr}
\bibinfo{author}{\bibfnamefont{H.~B.} \bibnamefont{Meyer}},
  \bibinfo{journal}{JHEP} \textbf{\bibinfo{volume}{03}}, \bibinfo{pages}{064}
  (\bibinfo{year}{2005}), \eprint{hep-lat/0412021}.

\bibitem[{\citenamefont{Lucini et~al.}(2005)\citenamefont{Lucini, Teper, and
  Wenger}}]{Lucini:2005vg}
\bibinfo{author}{\bibfnamefont{B.}~\bibnamefont{Lucini}},
  \bibinfo{author}{\bibfnamefont{M.}~\bibnamefont{Teper}}, \bibnamefont{and}
  \bibinfo{author}{\bibfnamefont{U.}~\bibnamefont{Wenger}},
  \bibinfo{journal}{JHEP} \textbf{\bibinfo{volume}{02}}, \bibinfo{pages}{033}
  (\bibinfo{year}{2005}), \eprint{hep-lat/0502003}.

\bibitem[{\citenamefont{Arvis}(1983)}]{Arvis:1983fp}
\bibinfo{author}{\bibfnamefont{J.~F.} \bibnamefont{Arvis}},
  \bibinfo{journal}{Phys. Lett.} \textbf{\bibinfo{volume}{B127}},
  \bibinfo{pages}{106} (\bibinfo{year}{1983}).

\bibitem[{\citenamefont{Bringoltz and Teper}(2006)}]{Bringoltz:2005xx}
\bibinfo{author}{\bibfnamefont{B.}~\bibnamefont{Bringoltz}} \bibnamefont{and}
  \bibinfo{author}{\bibfnamefont{M.}~\bibnamefont{Teper}},
  \bibinfo{journal}{Phys. Rev.} \textbf{\bibinfo{volume}{D73}},
  \bibinfo{pages}{014517} (\bibinfo{year}{2006}), \eprint{hep-lat/0508021}.

\bibitem[{\citenamefont{Yaffe and Svetitsky}(1982)}]{Yaffe:1982qf}
\bibinfo{author}{\bibfnamefont{L.~G.} \bibnamefont{Yaffe}} \bibnamefont{and}
  \bibinfo{author}{\bibfnamefont{B.}~\bibnamefont{Svetitsky}},
  \bibinfo{journal}{Phys. Rev.} \textbf{\bibinfo{volume}{D26}},
  \bibinfo{pages}{963} (\bibinfo{year}{1982}).

\bibitem[{\citenamefont{Beinlich et~al.}(1997)\citenamefont{Beinlich, Karsch,
  and Peikert}}]{Beinlich:1996xg}
\bibinfo{author}{\bibfnamefont{B.}~\bibnamefont{Beinlich}},
  \bibinfo{author}{\bibfnamefont{F.}~\bibnamefont{Karsch}}, \bibnamefont{and}
  \bibinfo{author}{\bibfnamefont{A.}~\bibnamefont{Peikert}},
  \bibinfo{journal}{Phys. Lett.} \textbf{\bibinfo{volume}{B390}},
  \bibinfo{pages}{268} (\bibinfo{year}{1997}), \eprint{hep-lat/9608141}.

\bibitem[{\citenamefont{Zwiebach}(2004)}]{Zwiebach:2004tj}
\bibinfo{author}{\bibfnamefont{B.}~\bibnamefont{Zwiebach}}
  (\bibinfo{year}{2004}), \bibinfo{note}{A First Course in String Theory, 
  Cambridge, UK: Univ. Pr. 558 p}.

\bibitem[{\citenamefont{Lucini et~al.}(2004)\citenamefont{Lucini, Teper, and
  Wenger}}]{Lucini:2003zr}
\bibinfo{author}{\bibfnamefont{B.}~\bibnamefont{Lucini}},
  \bibinfo{author}{\bibfnamefont{M.}~\bibnamefont{Teper}}, \bibnamefont{and}
  \bibinfo{author}{\bibfnamefont{U.}~\bibnamefont{Wenger}},
  \bibinfo{journal}{JHEP} \textbf{\bibinfo{volume}{01}}, \bibinfo{pages}{061}
  (\bibinfo{year}{2004}), \eprint{hep-lat/0307017}.

\bibitem[{\citenamefont{Meyer}(2009{\natexlab{b}})}]{Meyer:2008tr}
\bibinfo{author}{\bibfnamefont{H.~B.} \bibnamefont{Meyer}},
  \bibinfo{journal}{JHEP} \textbf{\bibinfo{volume}{01}}, \bibinfo{pages}{071}
  (\bibinfo{year}{2009}{\natexlab{b}}), \eprint{0808.3151}.

\bibitem[{\citenamefont{Lucini and Teper}(2001)}]{Lucini:2001ej}
\bibinfo{author}{\bibfnamefont{B.}~\bibnamefont{Lucini}} \bibnamefont{and}
  \bibinfo{author}{\bibfnamefont{M.}~\bibnamefont{Teper}},
  \bibinfo{journal}{JHEP} \textbf{\bibinfo{volume}{06}}, \bibinfo{pages}{050}
  (\bibinfo{year}{2001}), \eprint{hep-lat/0103027}.

\bibitem[{\citenamefont{Hagedorn}(1965)}]{Hagedorn:1965st}
\bibinfo{author}{\bibfnamefont{R.}~\bibnamefont{Hagedorn}},
  \bibinfo{journal}{Nuovo Cim. Suppl.} \textbf{\bibinfo{volume}{3}},
  \bibinfo{pages}{147} (\bibinfo{year}{1965}).

\bibitem[{\citenamefont{Gubser et~al.}(1996)\citenamefont{Gubser, Klebanov, and
  Peet}}]{Gubser:1996de}
\bibinfo{author}{\bibfnamefont{S.~S.} \bibnamefont{Gubser}},
  \bibinfo{author}{\bibfnamefont{I.~R.} \bibnamefont{Klebanov}},
  \bibnamefont{and} \bibinfo{author}{\bibfnamefont{A.~W.} \bibnamefont{Peet}},
  \bibinfo{journal}{Phys. Rev.} \textbf{\bibinfo{volume}{D54}},
  \bibinfo{pages}{3915} (\bibinfo{year}{1996}), \eprint{hep-th/9602135}.

\bibitem[{\citenamefont{Selem and Wilczek}(2006)}]{Selem:2006nd}
\bibinfo{author}{\bibfnamefont{A.}~\bibnamefont{Selem}} \bibnamefont{and}
  \bibinfo{author}{\bibfnamefont{F.}~\bibnamefont{Wilczek}}
  (\bibinfo{year}{2006}), \eprint{hep-ph/0602128}.

\end{thebibliography}

\end{document}